\documentclass[aps,prd,a4paper,onecolumn,amsmath,showpacs,superscriptaddress,nofootinbib,preprintnumbers,notitlepage]{revtex4-1}
 
\usepackage{verbatim}
\usepackage[T1]{fontenc}
\usepackage[utf8]{inputenc}
\usepackage[american]{babel}
\usepackage{epsfig}
\usepackage{array}
\usepackage{graphicx,subcaption,caption}
\usepackage{pgfplots}
\usepackage{booktabs}
\usepackage{multirow}
\usepackage{dcolumn}
\usepackage{amsmath}
\usepackage{mathtools}
\usepackage{amsfonts}
\usepackage{amssymb}
\usepackage{epstopdf}
\usepackage{bm}
\usepackage{siunitx}
\usepackage{braket}
\usepackage{enumitem}
\usepackage{soul}
\usepackage{color}
\usepackage{transparent}
\usepackage{pifont}
\usepackage[font={small}]{caption}
\definecolor{navyblue}{rgb}{0.0, 0.0, 0.5}
\definecolor{royalblue}{rgb}{0.25, 0.41, 0.88}
\definecolor{cadmiumgreen}{rgb}{0.0, 0.42, 0.24}
\definecolor{blue-violet}{rgb}{0.54, 0.17, 0.89}
\definecolor{darkviolet}{rgb}{0.58, 0.0, 0.83}
\definecolor{orange(colorwheel)}{rgb}{1.0, 0.5, 0.0}

\usepackage{hyperref}
\hypersetup{
    colorlinks=true, % false: boxed links; true: colored links
    linkcolor=royalblue, % color of internal links (change box color with linkbordercolor)
    citecolor=magenta}

\newcommand\be{\begin{equation}}
\newcommand\ee{\end{equation}}
\newcommand\bea{\begin{eqnarray}}
\newcommand\eea{\end{eqnarray}}

\newcommand\ie{{\it i.e.}~}

\usepackage{booktabs}
\usepackage{multirow}
\usepackage{dcolumn}
\usepackage{colortbl}

\definecolor{magenta(process)}{rgb}{1.0, 0.0, 0.56}

\definecolor{darkspringgreen}{rgb}{0.09, 0.45, 0.27}

\definecolor{royalblue(web)}{rgb}{0.25, 0.41, 0.88}

\begin{document}

\title{Barrow entropies in black hole thermodynamics}

\author{Salvatore Capozziello}
\email{capozziello@na.infn.it}
\affiliation{Dipartimento di Fisica "E. Pancini",Universit\`a degli Studi di Napoli Federico II", Via Cinthia, I-80126, Napoli, Italy}
\affiliation{Istituto Nazionale di Fisica Nucleare (INFN), sez. di Napoli, Via Cinthia 9, I-80126 Napoli, Italy}
\affiliation{Scuola Superiore Meridionale, Largo S. Marcellino 10, I-80138, Napoli, Italy}

\author{Mehdi Shokri}
\email{mehdishokriphysics@gmail.com}
\affiliation{Canadian Quantum Research Center, 204-3002 32 Ave, Vernon, BC V1T 2L7, Canada}
\affiliation{School of Physics, Damghan University, P. O. Box 3671641167, Damghan, Iran}

%\preprint{}
\begin{abstract}
We study the thermodynamic features of static, spherically-symmetric Schwarzschild black holes adopting different types of Barrow entropy. Specifically, in addition to the standard Barrow entropy,  we consider a logarithmic-corrected type of this entropy by taking into account some loop quantum gravity effects. Moreover, we investigate the black hole thermodynamics from the viewpoint of Barrow entropy in  presence of non-extensivity  effects coming from the Tsallis statistics.  Finally, we  compare  the results obtained for  different Barrow-based entropies.
%\\\\
%\\{\bf PACS:} 04.50.Kd; 98.80.Cq.
%\\{\bf Keywords}: Modified Theories of Gravity; Non-metricity; Cosmic Inflation.
\end{abstract}
\date{\today}
\maketitle
%\tableofcontents
\section{Introduction}
The theory of black holes in general relativity (GR) nicely reveals a fundamental connection between gravitation, thermodynamics and quantum mechanics \cite{Bardeen:1973gs,Bekenstein:1973ur,Hawking:1975vcx}. From a purely classical thermodynamics perspective, black holes act like a perfect absorber, without any thermal emission, with an absolute zero temperature. However, Hawking showed that, in quantum theory, black holes emit thermal radiation, the so-called {\it Hawking radiation} \cite{Hawking:1974rv,Hawking:1975vcx}. Moreover, he confirmed Bekenstein's conjecture \cite{Bekenstein:1972tm,Bekenstein:1973ur} by introducing the constant of proportionality $1/4$ between entropy and horizon area of a black hole in the context of the Bekenstein–Hawking (BH) formula $S=A/4$. This relationship was extended to the area of any surface, not necessarily the horizon, by the Ryu-Takayanagi formula, which relates the entanglement entropy of a boundary conformal field theory to a specific surface in its dual gravitational theory \cite{Ryu:2006ef}. The mentioned quantum effects allow us to present an interpretation of the laws of black hole mechanics as
physically corresponding to the ordinary laws of thermodynamics \cite{Wald:1999vt, Page:2004xp, Carlip:2014pma,Almheiri:2020cfm}. The discovered relationship between gravitation and thermodynamics was later developed by the "gravity-thermodynamics" conjecture \cite{Padmanabhan:2003gd,Padmanabhan:2009vy}, which states that Einstein's gravity may
arise from the laws of thermodynamics applied to spacetime itself \cite{Jacobson:1995ab}. Hence, by using the first law
of thermodynamics to the apparent horizon of the universe, the extracted Friedmann equations may lead to some interesting consequences on the cosmological scales \cite{Padmanabhan:2012ik}. 

Despite the noteworthy consequences of the BH formula, we are interested in considering some extensions to the BH entropy arising from non-extensive generalization of the horizon degrees of freedom (DOF) or quantum gravitational corrections of the black hole surface. Such non-Gaussian statistics provide a natural framework to study semiclassical effects in the context of  generalized uncertainty principle (GUP). 
In \cite{Tsallis:1987eu}, Tsallis introduced a non-extensive generalization of Boltzmann-Gibbs (BG) thermodynamics, parameterized by the index $q$, in the context of Tsallis statistics. The non-extensivity parameter $q$ refers to super-extensivity and sub-extensivity when
$q<1$ and $q>1$, respectively \cite{Tsallis:2002tp}. In fact, rare events occur for $0\leq q<1$ and frequent events occur for $q>1$ \cite{Tsallis:1998as,NIVEN2004444}, pointing out to the stretching or compressing the entropy curve to lower or higher maximum entropy positions. The consequences of these non-extensivity effects on the thermodynamic parameters have been discussed in \cite{Abe:2001ss}. This statistics shows a deviation from extensivity by $|q-1|$ in which $q\rightarrow1$ depicts the standard BG theory. As a direct consequence of the $q$-generalized entropy, the non-extensivity effects on the thermodynamic parameters have been argued in \cite{Abe:2001ss}. Moreover, Tsallis entropy can affect the Bose-Einstein distribution due to the maximal entropy principle \cite{Chen:2002hx}. Apart from the statistical properties, Tsallis entropy has been applied to investigate a wide range of gravitational and cosmological scenarios such as dark energy (DE) \cite{Barboza:2014yfe,Nunes:2015xsa,Zadeh:2018poj,Ghaffari:2018wks,Tavayef:2018xwx, SayahianJahromi:2018irq, Saridakis:2018unr,Sheykhi:2018dpn, Lymperis:2018iuz, Sheykhi:2019bsh,Huang:2019hex,Aditya:2019bbk,DAgostino:2019wko,Mamon:2020wnh,Mohammadi:2021wde,Dheepika:2021fqv,Nojiri:2022dkr}, inflation theory \cite{Keskin:2023ngx,Odintsov:2023vpj,Teimoori:2023hpv}, the physics of black holes \cite{deOliveira:2005jy,Tsallis:2012js,Komatsu:2013qia,Mejrhit:2020dpo,Abreu:2020wbz,Petridis:2023gqz,Shokri:2024elp}, gravitational waves \cite{Jizba:2024klq}, the cosmic microwave background (CMB) \cite{Bernui:2005hq,Bernui:2007wj} and modified Newtonian dynamics (MOND) \cite{Abreu:2014dna,Abreu:2018pua}. As a relativistic generalization of BG statistics, the Kaniadakis statistics \cite{Kana}, also called $\kappa$-statistics, deals with a non-exponential distribution function parameterized by the index $-1<\kappa<1$ usually fixed through the theory under consideration. Also, the $\kappa$-statistics naturally emerges from the special relativity perspective as shown in \cite{Kaniadakis:2002zz}. Analogous to Tsallis entropy, a broad spectrum of literature has been dedicated to investigating the constructive effects of relativistic corrections added to the BG statistics in the context of $\kappa$-statistics. For instance, see some applications in holographic dark energy (HDE) \cite{Drepanou:2021jiv,Hernandez-Almada:2021aiw,P:2022amn}, cosmic inflation  \cite{Nojiri:2022dkr,Odintsov:2023vpj,Lambiase:2023ryq}, black holes \cite{Abreu:2021avp,Abreu:2021kwu, Cimidiker:2023kle} and other cosmological situations  \cite{Luciano:2022eio}.

In addition to the above-mentioned statistics, some non-Gaussian statistics deal with entropies, which are modified by logarithmic and power-law terms coming from quantum gravitational effects in the physics of black holes. The existence of a logarithmic correction to the BH entropy is secured through different perspectives  such as
conical singularity and entanglement entropy \cite{Solodukhin:1994yz,Solodukhin:2011gn}, Euclidean action method \cite{Fursaev:1994te,Sen:2012dw,El-Menoufi:2015cqw,El-Menoufi:2017kew}, conformal anomaly \cite{Cai:2009ua}, Cardy formula \cite{Carlip:2000nv}, quantum tunneling \cite{Banerjee:2008cf,Banerjee:2008ry}, quantum geometry \cite{Kaul:2000kf}, Noether charge \cite{Aros:2010jb} and non-locality of quantum gravity \cite{Xiao:2021zly}. Finally, this correction could have a prominent role in contributing to avoid black hole singularity in the framework recently discussed in \cite{Capozziello:2024ucm}.

As a general consideration, taking into account self-gravity and backreaction effects to the metric of the black hole leads to a modification of the Hawking temperature and then of black hole entropy. The logarithmic-corrected entropy is described by two dimensionless constants $\alpha$ and $\beta$, which are usually model-dependent and not fixed by some loop quantum gravity (LQG) consideration. Adding a logarithmic-corrected term to the BH entropy provides a suitable framework to revisit  gravitational and cosmological models such as HDE models, inflation theory and black hole thermodynamics \cite{Jamil:2010xq,Cai:2010zw,Abreu:2020wbz,Abreu:2020dyu}. Recently, Barrow \cite{Barrow:2020tzx}  proposed a new class of non-Gaussian statistics by considering a fractal structure on the black hole surface coming from quantum gravitational effects. This power-law correction is valid for  static, spherically symmetric Schwarzschild black holes by setting the necessary conditions to have a finite volume while the surface area tends to infinity in the limit of increasing intricacy on arbitrarily small scales. The Barrow entropy is parameterized by the index $0\leq\Delta\leq1$, which depicts the corresponding quantum gravity effects. In the light of  "gravity-thermodynamics" conjecture, different cosmological phenomena have been studied from the Barrow statistics perspective using  modified dynamical equations, see e.g.  \cite{Saridakis:2020lrg}. As an important application, the holography principle \cite{tHooft:1993dmi} is studied from the Barrow entropy viewpoint in order to describe the late-time accelerating phase of the universe or DE through the Barrow holographic dark energy (BHDE) models \cite{Saridakis:2020zol,Dabrowski:2020atl,Huang:2021zgj,Nojiri:2021jxf,Luciano:2022viz,Chanda:2022tpk,Ghaffari:2022skp}. Moreover, the early-time inflationary epoch can be understood due to the quantum gravity effects, coming from Barrow entropy, inserted into  cosmological models \cite{Maity:2022gdy,Luciano:2023roh,Saha:2024dhr,Ghaffari:2022skp}. Since the Barrow statistics was first inspired by black hole physics, some recent works have been accomplished to clarify different physical aspects of black holes in the context of Barrow entropy \cite{Abreu:2020dyu,Abreu:2020wbz,Abreu:2021kwu,Abreu:2022pil,Jawad:2022lww,Wang:2022hun,
Abreu:2022pil}.  

In Refs. \cite{Abreu:2020wbz,Abreu:2020dyu}, the authors examined the equipartition law and the heat capacity of static, spherically-symmetric Schwarzschild black holes from the viewpoint of different Barrow entropies. Here, we attempt to provide a complete study of the thermal properties of the black hole in the context of Barrow-type statistics by revising the existing results and also studying other crucial thermodynamic quantities, in particular,  the black hole lifetime, in order to find the constructive effects of the adopted Barrow statistics. First, we study observable thermodynamic quantities in the context of standard Barrow statistics. Then, we consider Barrow logarithmic-corrected entropy by taking into account some quantum and thermal fluctuations, coming from LQG, on the black hole metric. Moreover, we show how  thermodynamic observables behave in  presence of non-extensivity  effects upon the geometry of a Barrow black hole. To achieve the purposes of the paper, we arrange the paper as follows. In Section \ref{s2}, we present black hole thermodynamics in the standard Barrow statistics. In Section \ref{s3}, we pursue the issue in Barrow entropy improved with a logarithmic correction term coming from LQG. Then,in Section \ref{s4}, we examine  thermal properties of the Schwarzschild black hole in a Barrow-based entropy with non-extensivity  effects coming from the Tsallis  statistics . 
Conclusions  are drawn in Section \ref{s5}. In the following, we choose the physical units $c=\hbar=k_B=1.$

\section{The Barrow entropy}\label{s2}
In 1904, Helge von Koch \cite{Koch} introduced one of the earliest fractals, called the "Koch snowflake", in which a 2-dimensional object can be constructed iteratively in a sequence of stages with a finite area and infinite perimeter. A 3-dimensional version of such structures can be found in the Sierpinski Gasket \cite{Sierpinski} and the Menger Sponge \cite{Menger}, which provide finite volume and infinite surface area. Inspired by the von Koch idea, John Barrow \cite{Barrow:2020tzx} proposed that quantum gravitational effects can produce some intricate and fractal structures on the static, spherically symmetric Schwarzschild black hole surface tending to infinity while the volume is finite. To show this, we suppose $N$ hierarchically smaller spheres, with the radius $\lambda$ times smaller than the original sphere, touching spheres around the black hole event horizon. By considering the hierarchy of radii $r_{n+1}=\lambda r_n$, the  volume and the area of the black hole, after an infinite number of steps,  are given by
\begin{equation}
V_{\infty}=\frac{4\pi}{3}R^{3}_g \sum\limits_{n=0}^\infty (N\lambda^3)^n,\hspace{1cm}A_{\infty}=4\pi R^{2}_g \sum\limits_{n=0}^\infty (N\lambda^2)^n,
\label{a1}    
\end{equation}
where $R_g=2GM$ is  the Schwarzschild radius of the black hole, $G$ is the Newton constant and $M$ is the black hole mass. Then, it is easy to see that in order to have a finite volume and an infinite area, in the limit $N\rightarrow\infty$, we need the condition $\lambda^{-2}<N<\lambda^{-3}$. In the Barrow statistics, the BH entropy modifies as \cite{Barrow:2020tzx}
\begin{equation}
S_{\text{B}}=\bigg(\frac{A_g}{A_{pl}}\bigg)^{1+\frac{\Delta}{2}},
\label{a2}   
\end{equation}
where the subscript $\text{B}$ denotes the Barrow entropy. Here, $A_g=4\pi R^2_{g}$ is the surface area of a Schwarzschild black hole and $A_{pl}\approx4G$ is the Planck area. The new index $0\leq\Delta\leq1$ reflects quantum gravity effects so that $\Delta=0$ and $\Delta=1$ are for smooth (BH case) and the most fractal structures, respectively. Note that, although the index $\Delta$ is responsible for quantum gravity corrections added to the Barrow entropy, it does not contain any quantum parameter. By plugging the surface area of a Schwarzschild black hole into Eq.(\ref{a2}), we can rewrite the  Barrorw entropy as follows
\begin{equation}
S_{\text{B}}=(4\pi G)^{1+\frac{\Delta}{2}}M^{2+\Delta}.
\label{a3}   
\end{equation}
Now that we know the black hole entropy, let us study other observable thermodynamic quantities. By using the definition $\frac{1}{T}=\frac{\partial{S}}{\partial{M}}$, the black hole temperature,  in the Barrow entropy, can be found as
\begin{equation}
T=\frac{1}{(2+\Delta)(4\pi G)^{1+\frac{\Delta}{2}}}\frac{1}{M^{1+\Delta}}.
\label{a4}   
\end{equation} 
Substituting $M$ from Eq.(\ref{a4}) into Eq.(\ref{a3}), the Barrow entropy, in terms of the temperature $T$, takes the form 
\begin{equation}
S_\text{B}=\frac{1}{\Big[(2+\Delta)\sqrt{4\pi G}\Big]^{{\frac{2+\Delta}{1+\Delta}}}}\frac{1}{T^{\frac{2+\Delta}{1+\Delta}}}.
\label{a5}
\end{equation}
Moreover, we find that the number of DOF, $N=4S$, in the context of Barrow statistics, is given by 
\begin{equation}
N=4(4\pi G)^{1+\frac{\Delta}{2}}M^{2+\Delta}.
\label{a6}    
\end{equation}
By combining Eqs.(\ref{a4}) and (\ref{a6}), the modified equipartition law takes the following form
\begin{equation}
M=\frac{1}{2}\Big(1+\frac{\Delta}{2}\Big)NT.
\label{a7}    
\end{equation}
It is easy to see that in the limit $\Delta\rightarrow0$, the standard equipartition law is recovered. In other words, a Schwarzschild black hole in the BH statistics, without any quantum gravity effects, obeys the standard equipartition law $M=\frac{1}{2}NT$. This means each DOF of the black hole contributes by the same amount $\frac{1}{2}$. From the viewpoint of Barrow's statistics, the black hole satisfies the modified equipartition law (\ref{a7}) in which DOF of the black hole contribute by the same amount $\frac{1}{2}(1+\frac{\Delta}{2})$, depending only on the Barrow factor $\Delta$. Therefore, a Schwarzschild black hole fulfills the equipartition theorem even in the presence of quantum gravity effects coming from Barrow's entropy. As an important thermodynamic parameter, the heat capacity of the black hole at constant volume, $C_V=T(\frac{\partial{S}}{\partial{T}})_V$, can be written as 
\begin{equation}
C_V=-(4\pi G)^{1+\frac{\Delta}{2}}\Big(\frac{2+\Delta}{1+\Delta}\Big)M^{2+\Delta},
\label{a8}    
\end{equation}
which shows an unstable Schwarzschild black hole when $\Delta\rightarrow0$. To have a stable black hole with $C_V>0$, we are required to consider $\frac{2+\Delta}{1+\Delta}<0$, which leads to the condition $-2<\Delta<-1$. Due to the inconsistency of this condition with the theoretically allowed values of $\Delta$, the Schwarzschild black hole is unstable even in the presence of quantum gravitational effects of the back hole surface. Consequently, the black hole evaporates by losing the energy and mass of emitted particles throughout the Hawking radiation process. Despite the instability of the Schwarzschild black hole, Barrow's correction extends the lifetime of the black hole through a non-zero Barrow index $\Delta$. To show this, we begin with the Boltzmann radiation law, which is modified in the context of Barrow statistics as \cite{Petridis:2023gqz}
\begin{equation}
J_\text{B}=-\frac{\frac{dE}{dt}}{A_g^{1+\frac{\Delta}{2}}}=\sigma T^4,
\label{a9}    
\end{equation}
where $\sigma$ is the Stefan-Boltzmann constant. Plugging the Schwarzschild black hole area and Eq.(\ref{a4}) into the above relation and then using $dE=dM$, the time evolution of black hole mass is obtained as
\begin{equation}
M=\Big(M_0^{3(1+\Delta)}-3\gamma(1+\Delta)t\Big)^{\frac{1}{3(1+\Delta)}},
\label{a10}    
\end{equation}
where $\gamma=(\frac{1}{16\pi^3G^2})^{1+\frac{\Delta}{2}}\frac{\sigma}{(2+\Delta)^4}$ and $M_0$ denotes the initial mass of black hole. After a full dissipation $M=0$, we find the black hole lifetime as
\begin{equation}
t=\frac{1}{3\gamma(1+\Delta)}M_0^{3(1+\Delta)},
\label{a11}    
\end{equation}
which reduces to $t\propto M_0^3$ in the limit $\Delta\rightarrow0$.
Considering the Barrow black hole as a closed thermodynamic system, the Helmholtz free energy $F=U-TS$, at a constant temperature, is given by
\begin{equation}
F=\frac{1+\Delta}{(2+\Delta)^{\frac{2+\Delta}{1+\Delta}}}\frac{1}{(\epsilon T)^{\frac{1}{1+\Delta}}},
\label{a12}     
\end{equation}
where $\epsilon=(4\pi G)^{1+\frac{\Delta}{2}}$. Thus, the corresponding partition function can be obtained as follows
\begin{equation}
Z=\exp{\bigg[-\frac{1+\Delta}{(2+\Delta)^{\frac{2+\Delta}{1+\Delta}}}\frac{1}{(\sqrt{4\pi G}T)^{\frac{2+\Delta}{1+\Delta}}}\bigg]}.
\label{a13}    
\end{equation}
\begin{table}
\centering
\captionsetup{justification=centering}
\begin{tabular}{c|c|c}
   \toprule\toprule

   \textbf{Object} & \textbf{Mass}& \textbf{Radius}\\
  \hline
     Supermassive black hole & $10^6 - 10^{10}$ $M_\odot$& $0.001-400$ AU\\
    \hline
Intermediate-mass black hole  &  $10^3 - 10^{5}$ $M_\odot$& $10^3$ km\\
    \hline
      Stellar-mass black hole& $1 - 10^{12}$ $M_\odot$ & 30 km\\
    \hline
 Micro black hole&  $10^{-18} - 10^{-38}$ $M_\odot$& Up to 0.1 mm \\
    \hline\hline
\end{tabular}
 \caption{Black hole classification by Schwarzschild radius with $M_\odot=1.9891\times10^{30}$kg,  AU$=149597870.7$km.}
  \label{tab1}
\end{table}
In Table \ref{tab1}, black holes are classified based on their Schwarzschild radius or equivalently by their mass. Besides the two well-known black holes, i.e. the stellar-mass black holes and the
supermassive black hole, for which we have observational evidences, two
hypothetical objects (not yet observed), intermediate-mass black hole and micro-black holes, which could have been formed shortly after the Big Bang, are also listed in the table. 

\begin{figure*}[!hbtp]
     \centering	\includegraphics[width=0.43\textwidth,keepaspectratio]{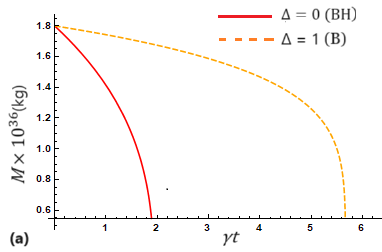}
\includegraphics[width=0.43\textwidth,keepaspectratio]{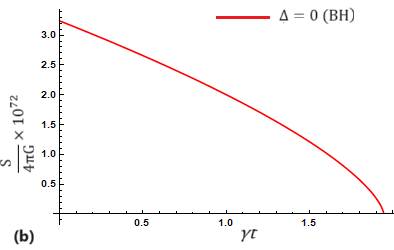}\vspace{0.5cm}
  \includegraphics[width=0.43\textwidth,keepaspectratio]{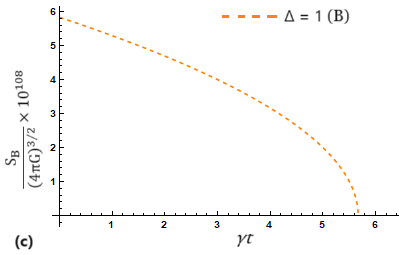}
\includegraphics[width=0.43\textwidth,keepaspectratio]{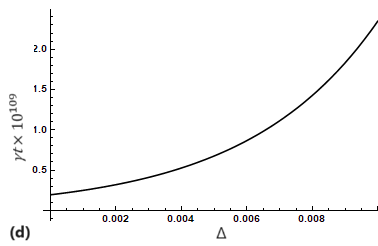}\vspace{0.5cm}

\includegraphics[width=0.43\textwidth,keepaspectratio]{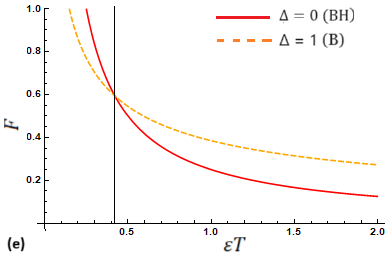}
\includegraphics[width=0.43\textwidth,keepaspectratio]{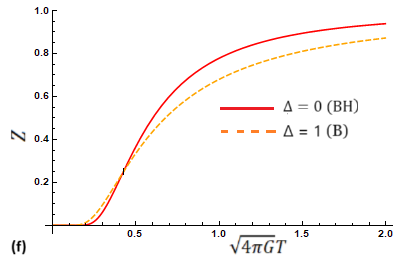}\vspace{0.5cm}
\caption{(a) Time evolution of the mass given in Eq. (\ref{a10}), (b) BH entropy $\Delta=0$ and (c) Barrow entropy in maximal deformation $\Delta=1$ (\ref{a3}). (d) The black hole lifetime versus the Barrow index $\Delta$ (\ref{a11}). (e) Helmholtz free energy of the Barrow black hole versus its temperature (\ref{a12}). (f) Partition function of the Barrow black hole versus its temperature (\ref{a13}). Panels (a) to (d) are drawn for the Schwarzschild black hole with an initial mass equal to the mass of the supermassive black hole SgrA$^*$, $M_0\sim10^{6}M_\odot$.}
\label{fig1}
\end{figure*}

Let us now study the Barrow index effects on the thermal features of a Schwarzschild black hole by drawing the appropriate plots. In Fig.\ref{fig1}, we present the time evolution of mass and entropy of the Schwarzschild black hole, with an initial mass equal to the mass of the supermassive black hole SgrA$^*$, $M_0\sim10^{6}M_\odot$, in the presence of a non-zero Barrow index $\Delta$. Panel (a) displays that the Barrow black hole ($\Delta\neq0$) loses mass slower than the BH black hole ($\Delta=0$) during the Hawking radiation process. Thus, the evaporation of Barrow-Schwarzschild black hole takes a longer time than the BH black hole. Panel (c) discloses that the entropy of the Schwarzschild black hole, in the presence of a non-zero Barrow index, approaches zero slower than its BH counterpart shown in panel (b). Moreover, panel (d) presents how the black hole lifetime behaves by varying the value of the Barrow index $\Delta$. We can see that the Schwarzschild black hole lifetime is extended by considering a deviation from the smooth case $\Delta=0$. In panel (e), we plot the Helmholtz free energy of a Barrow black hole versus its temperature. By considering a non-zero Barrow index, the Helmholtz free energy is always positive without showing any phase transition. This result is compatible with what we understood from the heat capacity (\ref{a8}), which says the Schwarzschild black hole in the context of the Barrow entropy is still unstable, similar to the case of BH entropy  ($\Delta=0$). Also, the panel reveals that the Helmholtz free energy for the most fractal structures ($\Delta=1$) equals its value for smooth structures ($\Delta=0$) at $T\sim0.42$. Moreover, the partition function associated with the Barrow black hole versus its temperature is drawn in panel (f). As we see, the general behavior of the partition function in  presence of a non-zero Barrow index is almost similar to the case of the BH black hole described by the canonical ensemble with state probability distribution in the form of BG statistical mechanics. In other words, Barrow's macroscopic modification of the BH entropy does not hurt the general microscopic properties of a standard Schwarzschild black hole ($\Delta=0$).
\section{The logarithmic-corrected Barrow entropy}\label{s3}
In this section, we first review the thermal properties of the Schwarzschild black hole in  presence of a logarithmic correction  added to the BH entropy. Then, we extend the analysis to the Barrow version of the Schwarzschild black hole.
\subsection{The Bekenstein–Hawking black hole}
By considering self-gravity and backreaction effects in the tunneling formalism of the Hawking radiation, the BH area law modifies as follows \cite{Banerjee:2008ry,Banerjee:2008cf}
\begin{equation}
S_\text{LC}=\bigg(\frac{A_g}{A_{pl}}\bigg)+\alpha\ln\bigg(\frac{A_g}{A_{pl}}\bigg)+\beta,
\label{b1}    
\end{equation}
where the subscript $\text{LC}$ indicates the logarithmic-corrected statistics. Moreover, $\alpha$ and $\beta$ are dimensionless constants with model-dependent values \cite{Gour:2002pj}. Inserting the Schwarzschild black hole surface area into the above equation, we can rewrite the black hole entropy  as
\begin{equation}
S_\text{LC}=\big(4\pi GM^2\big)+\alpha\ln\big(4\pi GM^2\big)+\beta,
\label{b2}
\end{equation}
with the corresponding temperature 
\begin{equation}
T=\frac{1}{2}\frac{M}{(\alpha+4\pi GM^2)}.
\label{b3}
\end{equation}
By solving the above equation, %Combining Eqs.(\ref{b2}) and (\ref{b3}), 
the black hole mass in terms of temperature takes the form
\begin{equation}
M_{\text{LC}}=\frac{1\pm\sqrt{1-64\pi G\alpha T^2}}{16\pi GT},
\label{b4}    
\end{equation}
%\begin{equation}
%S_{\text{LC}}=\frac{\big(1+\sqrt{1-64\pi G\alpha T^2}\big)^2}{64\pi GT^2}+\alpha\ln\Big[\frac{\big(1+\sqrt{1-64\pi G\alpha T^2}\big)^2}{64\pi GT^2}\Big]+\beta.
%\label{b4}    
%\end{equation}
where the positive sign guarantees the positiveness of the black hole mass while the negative sign leads to a negative mass for $\sqrt{1-64\pi G\alpha T^2}>1$. Now, let us find how equipartition law is modified by adding a logarithmic term to the BH statistics. The number of DOF $N=4S$ is found to be \begin{equation}
N=4\Big[\big(4\pi GM^2\big)+\alpha\ln\big(4\pi GM^2\big)+\beta\Big].
\label{b5}    
\end{equation}
Then using Eq.(\ref{b3}) and solving the corresponding equation, we have
\begin{equation}
M=2\alpha\big(1+W_k(X)\big)T\hspace{0.5cm}\text{where}\hspace{0.5cm}X=\frac{e^{\frac{N-4\beta}{4\alpha}}}{\alpha},
\label{b6}    
\end{equation}
where $W_{k}$ is the Lambert function with two important branches $k=0, -1$, returning real values for real input. By considering the asymptotic expansion of the Lambert function for large values of $X$ and keeping only the first term of the expansion, the above equation reduces to
\begin{equation}
M\simeq2\Big(\alpha[1-\ln(\alpha)]-\beta+\frac{N}{4}\Big)T.
\label{b7}    
\end{equation}
It is easy to see that when $\alpha$ and $\beta$ tend to zero, the standard equipartition law, $M=\frac{1}{2}NT$, is recovered. Note that by setting the condition $X\gg1$, one can obtain the constraint 
\begin{equation}
X\gg1\hspace{0.3cm}\longrightarrow\hspace{0.3cm}\alpha\ll1\hspace{0.3cm}\hspace{0.3cm}\text{or}\hspace{0.3cm}\beta\ll\frac{N}{4}.
\label{b8}
\end{equation}
As an alternative analysis, we rewrite Eq.(\ref{b5}) as \cite {Abreu:2020dyu}
\begin{equation}
N\simeq4\Big[\big(4\pi GM^2\big)+\alpha\big(1-(4\pi GM^2)^{-1}\big)+\beta\Big].
\label{b9}
\end{equation}
Here, we have ignored the second and higher orders of the term $(4\pi GM^2)^{-1}$ after using the logarithmic expansion 
\begin{equation}
\ln(x)=-\ln(\frac{1}{x})=\sum\limits_{n=1}^\infty\frac{(x-1)^n}{nx^n}\hspace{0.5cm}\text{if}\hspace{0.5cm}Re(x)\geq\frac{1}{2}.
\label{b10}    
\end{equation}
Now, by combining Eqs.(\ref{b3}) and (\ref{b9}), we find the modified equipartition law as follows
\begin{equation}
M\simeq\bigg(\frac{N}{4}+\alpha-\beta\pm\sqrt{4\alpha+\Big(\frac{N}{4}-\alpha-\beta\Big)^2}\bigg)T,
\label{b11}    
\end{equation}
where the positive sign navigates us to the standard equipartition law when $\alpha=\beta=0$. In addition to the obtained observables, the heat capacity of the logarithmic-corrected BH black hole can be found by 
\begin{equation}
C_V=-\frac{2(\alpha+4\pi GM^2)^2}{4\pi GM^2-\alpha},
\label{b12}    
\end{equation}
which reduces to the heat capacity of an unstable Schwarzschild black hole when $\alpha\rightarrow0$. It is easy to see that the stability condition $C_V>0$ causes the constraint $\alpha>4\pi GM^2$ which is not compatible with the obtained condition (\ref{b8}). Hence, the Schwarzschild black hole studied in the presence of LQG effects is still an unstable black hole with negative heat capacity. Now, it is worth checking how  the logarithmic correction term affects the lifetime of an unstable Schwarzschild black hole. Let us start with the modified Boltzmann radiation law
\begin{equation}
J_\text{LC}=-\frac{\frac{dE}{dt}}{\mathcal{A}}=\sigma T^4,
\label{b13}    
\end{equation}
where $\mathcal{A}=A_g+\alpha\ln(A_g)+\beta$ and $\sigma$ is the Stefan-Boltzmann constant. Considering $dE=dM$ and then substituting the temperature and surface area of the black hole into the above relation, we have the following expression
\begin{equation}
\frac{dM}{dt}=-\frac{\sigma}{16}\frac{M^4\Big(16\pi G^2M^2+\alpha\ln\big(16\pi G^2M^2\big)+\beta\Big)}{(\alpha+4\pi GM^2)^4},
\label{b14}    
\end{equation}
which has no analytical solution. %\textcolor{blue}{We try to solve the integration by the numerical methods.} 
Therefore, we attempt to find an analytical expression for the black hole lifetime by neglecting the role of $\alpha$ with respect to the term $4\pi GM^2$ due to the instability condition $\alpha\ll4\pi GM^2$. Then, the black hole mass versus time $t$ behaves 
\begin{equation}
\Big(4\pi^2GM(16\pi G^2M^2-3\beta)+3\sqrt{(\pi\beta)^3}\arctan\big[4GM\sqrt{\frac{\pi}{\beta}}\big]\Big)\Bigg|^{M_{0}}_{M}\simeq\frac{3G\sigma}{4}t,
\label{b15}
\end{equation}
where $M_0$ denotes the black hole mass before starting the evaporation process. By setting a fully evaporation condition $M=0$, the black hole lifetime can be found as
\begin{equation}
t\simeq\frac{4}{3G\sigma}\Big(4\pi^2GM_0(16\pi G^2M_0^2-3\beta)+3\sqrt{(\pi\beta)^3}\arctan\big[4GM_0\sqrt{\frac{\pi}{\beta}}\big]\Big),
\label{b16}    
\end{equation}
which approaches the Schwarzschild black hole lifetime in the BH statistics $t\propto M_0^3$ when $\beta$ goes to zero. Besides the obtained quantities, the Helmholtz free energy of the logarithmic-corrected black hole is given by
\begin{equation}
F=\frac{1+\sqrt{1-64\pi G\alpha T^2}-32\pi GT^2\bigg(\alpha\Big(\ln\big[\frac{(1+\sqrt{1-64\pi G\alpha T^2})^2}{64\pi G T^2}\big]-1\Big)+\beta\bigg)}{32\pi GT},
\label{b17}    
\end{equation}
with an appropriate partition function
\begin{equation}
Z=\exp{\Bigg[-\frac{1+\sqrt{1-64\pi G\alpha T^2}-32\pi GT^2\bigg(\alpha\Big(\ln\big[\frac{(1+\sqrt{1-64\pi G\alpha T^2})^2}{64\pi G T^2}\big]-1\Big)+\beta\bigg)}{32\pi GT^2}}\Bigg].
\label{b18}    
\end{equation}
In the next subsection, we investigate the effects of the above logarithmic term on the thermodynamics of the Barrow black hole.
\subsection{The Barrow black hole}
Let us now  analyze the thermal characteristics of a Schwarzschild black hole when  quantum gravity effects are taken into account by adding a logarithmic-corrected term into Barrow entropy. We start with  logarithmic-corrected Barrow entropy 
\begin{equation}
S_\text{BLC}=\bigg(\frac{A_g}{A_{pl}}\bigg)^{1+\frac{\Delta}{2}}+\alpha(1+\frac{\Delta}{2})\ln\bigg(\frac{A_g}{A_{pl}}\bigg)+\beta,
\label{c1}    
\end{equation}
where the subscript $\text{BLC}$ implies to the  logarithmic-corrected Barrow entropy. Also, $\Delta$, $\alpha$ and $\beta$ are model parameters. Considering the Schwarzschild black hole and substituting its surface area, the above entropy takes the form
\begin{equation}
S_\text{BLC}=\big(4\pi GM^2\big)^{1+\frac{\Delta}{2}}+\alpha(1+\frac{\Delta}{2})\ln\big(4\pi GM^2\big)+\beta.
\label{c2}    
\end{equation}
Using the definition $\frac{1}{T}=\frac{\partial{S}}{\partial{M}}$, the black hole temperature is given by
\begin{equation}
T=\frac{M}{(2+\Delta)\Big(\alpha+(4\pi GM^2)^{1+\frac{\Delta}{2}}\Big)}.
\label{c3}
\end{equation}
It is easy to see that the above equation cannot be  analytically solved  to find an expression of the black hole mass in terms of its temperature. Therefore, by considering the allowed range of  Barrow index $0\leq\Delta\leq1$, the black hole mass, in the case of maximal deformation $\Delta=1$, is obtained as
\begin{eqnarray}
M_{\text{BLC($\Delta=1$)}}=\frac{1}{3}\Bigg[\frac{2^{\frac{1}{3}}}{\Big(-27(4\pi G)^{3}\alpha T^3+\sqrt{-4(4\pi G)^{\frac{9}{2}}T^3+729(4\pi G)^{6}\alpha^2 T^6}\Big)^{\frac{1}{3}}}+\hspace{4cm}\nonumber\\
+\frac{\Big(-27(4\pi G)^{3}\alpha T^3+\sqrt{-4(4\pi G)^{\frac{9}{2}}T^3+729(4\pi G)^{6}\alpha^2 T^6}\Big)^{\frac{1}{3}}}{2^{\frac{1}{3}}(4\pi G)^{\frac{3}{2}}T}\Bigg].\hspace{0.2cm}
\label{cn1}    
\end{eqnarray}
Notice that, for the smooth case $\Delta=0$, the result of the  logarithmic-corrected BH black hole (\ref{b4}) is recovered. For the number of DOF $N=4S$, we find
\begin{equation}
N=4\Big[\big(4\pi GM^2\big)^{1+\frac{\Delta}{2}}+\alpha(1+\frac{\Delta}{2})\ln\big(4\pi GM^2\big)+\beta\Big].
\label{c4}    
\end{equation}
Now using the relation (\ref{c3}), the above equation for the black hole mass can be solved as
\begin{equation}
M=\alpha(2+\Delta)\big(1+W_k(X)\big)T\hspace{0.5cm}\text{where}\hspace{0.5cm}X=\frac{e^{\frac{N-4\beta}{4\alpha}}}{\alpha}.
\label{c5}    
\end{equation}
Analogous to the previous section and under the condition (\ref{b8}), the above relation reduces to 
\begin{equation}
M\simeq(2+\Delta)\Big(\alpha[1-\ln(\alpha)]-\beta+\frac{N}{4}\Big)T.
\label{c6}    
\end{equation}
%\begin{figure*}[!hbtp]
 %    \centering	\includegraphics[width=0.43\textwidth,keepaspectratio]{.png}
%\includegraphics[width=0.43\textwidth,keepaspectratio]{.png}\vspace{0.5cm}
  %\includegraphics[width=0.43\textwidth,keepaspectratio]{.png}
%\includegraphics[width=0.43\textwidth,keepaspectratio]{.png}\vspace{0.5cm}
%\caption{Time evolution of the mass (a), BH logarithmic-corrected entropy $\Delta=0$ (b) and Barrow logarithmic-corrected entropy in maximal deformation $\Delta=1$ (c) of Schwarzschild black hole. The black hole lifetime versus the Barrow index $\Delta$ (d). All panels are drawn for the Schwarzschild black hole with an initial mass equal to the mass of the supermassive black hole SagA$^*$, $M_0\sim10^{-6}M_\odot$.}
%\label{fig2}
%\end{figure*}
Notice that the standard equipartition law is recovered when $\Delta$, $\alpha$ and $\beta$ tend to zero. As an alternative solution, we rewrite the relation (\ref{c4}) as
\begin{equation}
N\simeq4\Big[\big(4\pi GM^2\big)^{1+\frac{\Delta}{2}}+\alpha\Big(1-(4\pi GM^2)^{-1-\frac{\Delta}{2}}\Big)+\beta\Big],
\label{c7}    
\end{equation}
in which the higher orders of  $(4\pi GM^2)^{-1-\frac{\Delta}{2}}$ can be  neglected  applying the logarithmic expansion (\ref{b10}). Now, by mixing Eqs.(\ref{c3}) and (\ref{c7}), the modified equipartition law can be found as
\begin{equation}
M\simeq\frac{2+\Delta}{2}\bigg(\frac{N}{4}+\alpha-\beta\pm\sqrt{4\alpha+\Big(\frac{N}{4}-\alpha-\beta\Big)^2}\bigg)T,
\label{c8}    
\end{equation}
where, for the positive sign, the standard equipartition law is recovered  for $\Delta=\alpha=\beta=0$. Notice that both relations (\ref{c6}) and (\ref{c8}) reduce to their counterparts in the BH entropy (\ref{b7}) and (\ref{b11}) if $\Delta\rightarrow0$. Moreover, we find that the logarithmic-corrected Barrow black hole obeys the modified equipartition law (\ref{c6}), or equivalently (\ref{c8}), telling us that the DOF of the black hole contribute by the amount 
\begin{equation}
\begin{cases}
    \frac{2+\Delta}{N}\Big(\alpha[1-\ln(\alpha)]-\beta+\frac{N}{4}\Big)\\
     \text{or equivalently,}\\
     \frac{2+\Delta}{2N}\bigg(\frac{N}{4}+\alpha-\beta+\sqrt{4\alpha+\Big(\frac{N}{4}-\alpha-\beta\Big)^2}\bigg),
\end{cases}
\label{cc8}    
\end{equation}
depending on Barrow and logarithmic-corrected entropies factors \ie $\Delta$, $\alpha$, $\beta$ and the number of DOF $N$. As we see, a Schwarzschild black hole satisfies the
equipartition theorem when quantum gravity effects from Barrow and logarithmic-corrected entropies are considered spontaneously. Note that this result is also valid in the case of a logarithmic-corrected BH black hole when $\Delta\rightarrow0$. Moreover, the heat capacity of the logarithmic-corrected Barrow black hole is given by
\begin{equation}
C_V=-\frac{(2+\Delta)\Big(\alpha+(4\pi GM^2)^{1+\frac{\Delta}{2}}\Big)^{2}}{(1+\Delta)(4\pi GM^2)^{1+\frac{\Delta}{2}}-\alpha}.
\label{c9}    
\end{equation}
\begin{figure*}[!hbtp]
     \centering	\includegraphics[width=0.43\textwidth,keepaspectratio]{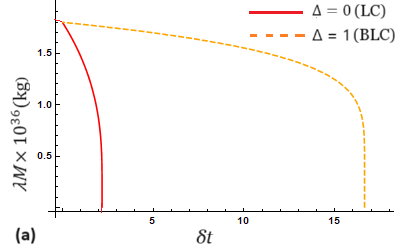}
\includegraphics[width=0.43\textwidth,keepaspectratio]{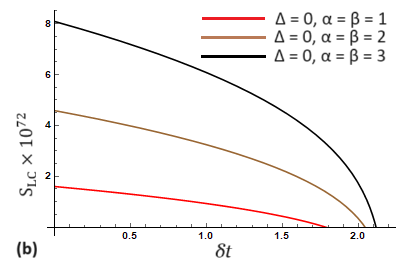}\vspace{0.5cm}
\includegraphics[width=0.43\textwidth,keepaspectratio]{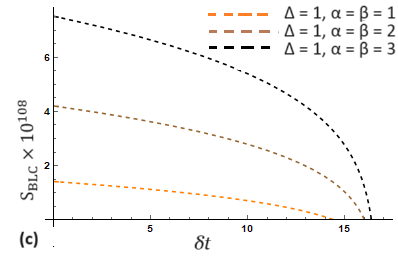}
\includegraphics[width=0.43\textwidth,keepaspectratio]{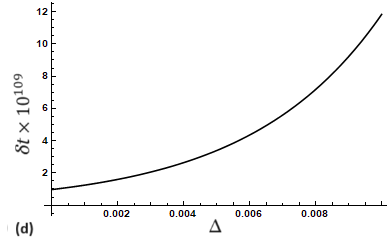}\vspace{0.5cm}
\includegraphics[width=0.43\textwidth,keepaspectratio]{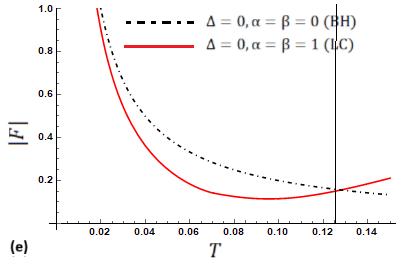}
\includegraphics[width=0.43\textwidth,keepaspectratio]{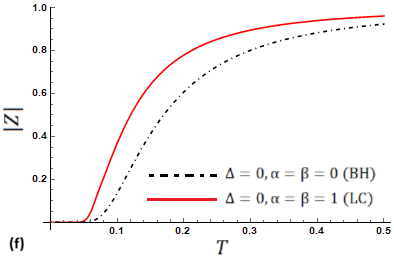}\vspace{0.5cm}
\includegraphics[width=0.43\textwidth,keepaspectratio]{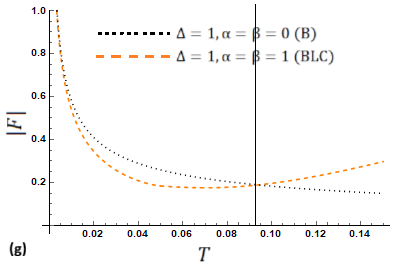}
\includegraphics[width=0.43\textwidth,keepaspectratio]{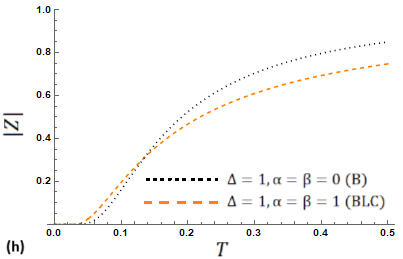}\vspace{0.5cm}
\caption{(a) Time evolution of the mass (\ref{c13}), (b) BH entropy $\Delta=0$ in the presence of a logarithmic correction term and (c) Barrow entropy in maximal deformation $\Delta=1$ in the presence of a logarithmic correction term (\ref{c2}). The black hole lifetime in the logarithmic-corrected Barrow statistics versus the Barrow index $\Delta$ (\ref{c14}). (e) Helmholtz free energy of the logarithmic-corrected BH black hole versus its
temperature (\ref{b17}). (f) Partition function of the logarithmic-corrected BH black hole versus its temperature (\ref{b18}). (g) Helmholtz free energy of the logarithmic-corrected Barrow black hole, in the case of maximal deformation $\Delta=1$, versus its
temperature (\ref{cn3}). (h) Partition function of the logarithmic-corrected Barrow black hole, in the case of maximal deformation $\Delta=1$, versus its temperature. Panels (a) to (d) are drawn for the Schwarzschild black hole with an initial mass proportional to the mass of the supermassive black hole SgrA$^*$, $\lambda M_0\sim10^{6}M_\odot$. Also, in panels (b), (c) and (e) to (h), we have used the assumption $G=1$.}
\label{fig2}
\end{figure*}
Note that the above expression reduces to the case of an unstable BH black hole when $\Delta$ and $\alpha$ approach zero. Now, it is worth checking out the stability condition of a Schwarzschild black hole introduced in the context of Barrow logarithmic-corrected statistics. By setting the stability condition $C_V>0$, we have two configurations as follow: 
\begin{equation}
\frac{(2+\Delta)}{(1+\Delta)(4\pi GM^2)^{1+\frac{\Delta}{2}}-\alpha}<0\hspace{0.5cm}\longrightarrow\hspace{0.5cm}
\begin{cases}
    (2+\Delta)>0  \hspace{0.2cm} \text{and} \hspace{0.2cm}(1+\Delta)(4\pi GM^2)^{1+\frac{\Delta}{2}}-\alpha<0,\\
    (2+\Delta)<0 \hspace{0.2cm} \text{and} \hspace{0.2cm} (1+\Delta)(4\pi GM^2)^{1+\frac{\Delta}{2}}-\alpha>0.
  \end{cases}
  \label{c10}
\end{equation}
Although the first configuration navigates us to some allowed values of $\Delta$, it shows values of $\alpha$ which are not in good agreement with the model constraint \ie $\alpha\ll(4\pi GM^2)^{1+\frac{\Delta}{2}}$. Moreover, the second set of conditions is not acceptable due to the inconsistency of $\Delta<-2$ with the theoretical prediction of the Barrow entropy which says $0\leq\Delta\leq1$. Hence, Barrow's black hole is unstable even by considering a logarithmic correction term. To study LQG effects on the lifetime of the Barrow black hole, we introduce the related Boltzmann radiation law 
\begin{equation}
J_\text{BLC}=-\frac{\frac{dE}{dt}}{\mathcal{A}}=\sigma T^4,
\label{c11}    
\end{equation}
where $\mathcal{A}=A_g^{1+\frac{\Delta}{2}}+\alpha(1+\frac{\Delta}{2})\ln(A_g)+\beta$ and $\sigma$ is the Stefan-Boltzmann constant. By substituting the black hole area and its temperature, we find 
\begin{equation}
\frac{dM}{dt}=-\frac{\sigma}{(2+\Delta)^4}\frac{M^4\Big((16\pi G^2M^2)^{1+\frac{\Delta}{2}}+\alpha(1+\frac{\Delta}{2})\ln\big(16\pi G^2M^2\big)+\beta\Big)}{\big(\alpha+(4\pi GM^2)^{1+\frac{\Delta}{2}}\big)^4}.
\label{c12}    
\end{equation}
In order to have an analytical solution, we neglect the role of $\alpha$ with respect to the term $(4\pi GM^2)^{1+\frac{\Delta}{2}}$ under the instability condition $\alpha\ll(4\pi GM^2)^{1+\frac{\Delta}{2}}$. Then, the time evolution of black hole mass is given by 
%\begin{equation}
%\frac{(4\pi G)^{4+2\Delta}M^{5+4\Delta}}{\beta(5+4\Delta)}{}_2 F_1\bigg(1,\frac{5+4\Delta}{2+\Delta};\frac{7+5\Delta}{2+\Delta};-\frac{(16\pi G^2M^2)^{1+\frac{\Delta}{2}}}{\beta}\bigg)\Bigg|^{M_{0}}_{M}\simeq\frac{\sigma}{(2+\Delta)^4}t,
%\label{}    
%\end{equation}
\begin{equation}
(\lambda M)^{5+4\Delta}{}_2 F_1\bigg(1,\frac{5+4\Delta}{2+\Delta};\frac{7+5\Delta}{2+\Delta};-(\lambda M)^{2+\Delta}\bigg)\Bigg|^{M_{0}}_{M}\simeq\delta t,
\label{c13}
\end{equation}
where $\lambda=\frac{\sqrt{16\pi G^2}}{\beta^{\frac{1}{2+\Delta}}}$ and $\delta=\frac{\sigma(5+4\Delta)(4G)^{1+2\Delta}}{\sqrt{\pi^3}\beta^{\frac{3(1+\Delta)}{2+\Delta}}(2+\Delta)^4}$. Also, ${}_2 F_1$ is the hypergeometric function and $M_0$ indicates the initial mass of the black hole. After a full evaporation $M=0$, the black hole lifetime can be calculated as
%\begin{equation}
%t\simeq\frac{(4\pi G)^{4+2\Delta}M_0^{5+4\Delta}(2+\Delta)^4}{\beta\sigma(5+4\Delta)}{}_2 F_1\bigg(1,\frac{5+4\Delta}{2+\Delta};\frac{7+5\Delta}{2+\Delta};-\frac{(16\pi G^2M_0^2)^{1+\frac{\Delta}{2}}}{\beta}\bigg),
%\label{}    
%\end{equation}
\begin{equation}
t\simeq\frac{1}{\delta}(\lambda M_0)^{5+4\Delta}{}_2 F_1\bigg(1,\frac{5+4\Delta}{2+\Delta};\frac{7+5\Delta}{2+\Delta};-(\lambda M_0)^{2+\Delta}\bigg),
\label{c14}    
\end{equation}
which reduces to the standard black hole lifetime $t\propto M_0^3$ if $\Delta$, $\alpha$ and $\beta$ tend to zero. Also, in the limit of $\Delta\rightarrow0$, the above expression approaches the BH black hole lifetime, in the presence of the logarithmic-corrected term, presented in Eq.(\ref{b16}). For the sake of completeness, we find the Helmholtz free energy of the  logarithmic-corrected Barrow black hole, in the case of maximal deformation $\Delta=1$, as follows
\begin{eqnarray}
F_{\Delta=1}=\frac{1}{9}\Bigg[\frac{2^{\frac{4}{3}}}{\Big(-27(4\pi G)^{3}\alpha T^3+\sqrt{-4(4\pi G)^{\frac{9}{2}}T^3+729(4\pi G)^{6}\alpha^2 T^6}\Big)^{\frac{1}{3}}}+\hspace{4cm}\nonumber\\+\frac{2^{\frac{2}{3}}\Big(-27(4\pi G)^{3}\alpha T^3+\sqrt{-4(4\pi G)^{\frac{9}{2}}T^3+729(4\pi G)^{6}\alpha^2 T^6}\Big)^{\frac{1}{3}}}{(4\pi G)^{\frac{3}{2}}T}-9T\times\hspace{2cm}\nonumber\\
\times\Bigg(\alpha\bigg(\ln\Big[\frac{\Big(2^{\frac{4}{3}}(4\pi G)^{\frac{3}{2}}T+\big(-54(4\pi G)^{3}\alpha T^3+2\sqrt{-4(4\pi G)^{\frac{9}{2}}T^3+729(4\pi G)^{6}\alpha^2 T^6}\big)^{\frac{2}{3}}\Big)^{3}}{216(4\pi G)^{3}T^3\Big(-27(4\pi G)^{3}\alpha T^3+\sqrt{-4(4\pi G)^{\frac{9}{2}}T^3+729(4\pi G)^{6}\alpha^2 T^6}\Big)}\Big]-1\bigg)+\beta\Bigg)\Bigg].
\label{cn3}    
\end{eqnarray}
%Also, the partition function related to the obtained Helmholtz free energy is given by
%\begin{eqnarray}
%Z=\exp\Bigg\{-\frac{1}{9T}\bigg[\frac{2^{\frac{4}{3}}}{\Big(-27(4\pi G)^{3}\alpha T^3+\sqrt{-4(4\pi G)^{\frac{9}{2}}T^3+729(4\pi G)^{6}\alpha^2 T^6}\Big)^{\frac{1}{3}}}+\hspace{4cm}\nonumber\\+\frac{2^{\frac{2}{3}}\Big(-27(4\pi G)^{3}\alpha T^3+\sqrt{-4(4\pi G)^{\frac{9}{2}}T^3+729(4\pi G)^{6}\alpha^2 T^6}\Big)^{\frac{1}{3}}}{(4\pi G)^{\frac{3}{2}}T}-9T\times\hspace{2cm}\nonumber\\
%\times\Bigg(\alpha\bigg(\ln\Big[\frac{\Big(2^{\frac{4}{3}}(4\pi G)^{\frac{3}{2}}T+\big(-54(4\pi G)^{3}\alpha T^3+2\sqrt{-4(4\pi G)^{\frac{9}{2}}T^3+729(4\pi G)^{6}\alpha^2 T^6}\big)^{\frac{2}{3}}\Big)^{3}}{216(4\pi G)^{3}T^3\Big(-27(4\pi G)^{3}\alpha T^3+\sqrt{-4(4\pi G)^{\frac{9}{2}}T^3+729(4\pi G)^{6}\alpha^2 T^6}\Big)}\Big]-1\bigg)+\beta\Bigg)\bigg]\Bigg\},
%\label{cn4}    
%\end{eqnarray}
Fig.\ref{fig2} shows the time evolution of mass and entropy of the Schwarzschild black hole in the context of the  logarithmic-corrected Barrow statistics, with an initial mass proportional to the mass of the supermassive black hole SgrA$^*$, $\lambda M_0\sim10^{6}M_\odot$. In panel (a), we see that evaporation of the  logarithmic-corrected Barrow black hole takes a longer time than that of the  logarithmic-corrected BH black hole. Thus, we have to consider self-gravity and backreaction effects along with the fractal structures on the black hole surface play a constructive role in delaying the evaporation of the black hole mass during the Hawking radiation process. This process is compatible with the result of panel (d), showing the lifetime of a  logarithmic-corrected Barrow black hole  extended by increasing the value of the Barrow index. Panels (b) and (c) present the entropy of the logarithmic-corrected versions of BH and Barrow black holes for different values of $\alpha$ and $\beta$. Due to the presence of a non-zero Barrow index, the entropy of the  logarithmic-corrected Barrow black hole goes to zero unhurriedly than the entropy of the logarithmic-corrected BH black hole. Also, panels (e) and (f)  show the Helmholtz free energy and the associated partition function of a logarithmic-corrected BH black hole versus its temperature for $\alpha=\beta=1$. Compared to a standard BH black hole, we realize that adding a logarithmic term into the BH entropy does not provide any constructive mechanism in order to get a stable black hole since the positive Helmholtz free energy does not show any phase transition. This is consistent with the result obtained from the heat capacity of the logarithmic-corrected BH black hole (\ref{b12}). Moreover, panel (e) tells us that, at $T\sim0.125$, the Helmholtz free energy of a logarithmic-corrected BH black hole equals its value for a standard BH black hole. In addition to the mentioned points, panel (f) reveals that loop quantum gravity effects on the BH entropy do not change the general microscopic properties of a BH Schwarzschild black hole. In the case of the logarithmic-corrected Barrow black hole, panels (g) and (h) present the behavior of its Helmholtz free energy and the corresponding partition function versus its temperature for $\alpha=\beta=1$. As we see, a Schwarzschild black hole is unstable even by taking into account self-gravity and backreaction effects to the black hole metric, coming from logarithmic-corrected statistics, along with quantum gravity effects on the black hole surface, coming from Barrow's statistics, due to the lack of phase transition in the behavior of Helmholtz free energy. This is, as we expect, from the heat capacity of a logarithmic-corrected Barrow black hole as shown in Eq.(\ref{c9}). Also, the temperature for the equality of Helmholtz free energy of Barrow black hole and its logarithmic-corrected version is $T\sim0.093$.

\section{The Barrow entropy with non-extensivity effects}\label{s4}
In the present section, we  examine the thermodynamic observables of the Schwarzschild black hole in  presence of non-extensivity effects added to the logarithmic-corrected version of BH and Barrow entropies. To perform this analysis, we define the non-extensive $q$-generalized statistics based on the non-additive Tsallis entropy \cite{Tsallis:1987eu}
\begin{equation}
S_\text{T}=\frac{1-\sum\limits_{i=1}^Wp_i^q}{q-1},
\label{d1}    
\end{equation}
where $p_i$, $W$ and $q$ refer to the probability of the existence of the system, total number of microstates and non-extensivity parameter, respectively. Also, the non-extensivity parameter $0\leq q<1$ and $q>1$ correspond to rare and frequent events, respectively. In the case of a microcaninical regime, where all microstates can be assumed by the same probability, Tsallis entropy reduces to 
\begin{equation}
S_\text{T}=\ln_qW=\frac{W^{1-q}-1}{1-q},
\label{d2}    
\end{equation}
where, the BG entropy, $S_\text{BG}=\ln W$, is recovered when $q\rightarrow1$. In the following, we discuss the black hole thermodynamics by considering some non-extensivity effects inserted into the geometry of BH and Barrow black holes which are contaminated by backreaction effects coming from LQG. 
\subsection{ The logarithmic-corrected Bekenstein-Hawking black hole}
For a non-extensive generalization of the  logarithmic-corrected BH statistics, we modify the entropy relation (\ref{b1}) as \cite{Abreu:2020wbz}
\begin{equation}
S_\text{TLC}=\bigg(\frac{A_g}{A_{pl}}\bigg)+\alpha\ln_q\bigg(\frac{A_g}{A_{pl}}\bigg)+\beta,
\label{d3}    
\end{equation}
where the subscript $\text{TLC}$ denotes the  logarithmic-corrected BH entropy in the context of  Tsallis statistics. Plugging the surface area of the Schwarzschild black hole and Tsallis entropy (\ref{d2}) into the above relation, we have
\begin{equation}
S_{\text{TLC}}=4\pi GM^2-\frac{\alpha\Big(1-(4\pi GM^2)^{1-q}\Big)}{1-q}+\beta,
\label{d4}    
\end{equation}
with the corresponding temperature
\begin{equation}
T=\frac{1}{8\pi GM+2\alpha(4\pi G)^{1-q}M^{1-2q}}.
\label{d5}    
\end{equation}
This equation can not be solved analytically, however, we are able to find an expression of the black hole mass versus the temperature for some values of the non-extensivity parameter $q$. For $q=1$, we trivially reproduce the result of the  logarithmic-corrected BH black hole presented in Eq.(\ref{b4}). For $q=2$, we find the following solutions for the black hole mass in terms of the temperature 
\begin{equation}
M_\text{TLC($q=2$)}=\frac{1}{2}\Bigg(\frac{1}{16\pi GT}-\sqrt{\mathcal{A}}\mp\sqrt{\frac{3}{4(8\pi G)^2T^2}-\mathcal{A}-\frac{1}{4(8\pi G)^3T^3\sqrt{\mathcal{A}}}}\Bigg),
\label{d6}    
\end{equation}
\begin{equation}
M_\text{TLC($q=2$)}=\frac{1}{2}\Bigg(\frac{1}{16\pi GT}+\sqrt{\mathcal{A}}\mp\sqrt{\frac{1}{4(8\pi G)^2T^2}-\mathcal{A}+\frac{1}{4(8\pi G)^3T^3\sqrt{\mathcal{A}}}}\Bigg),
\label{d7}    
\end{equation}
where 
\begin{equation}
\mathcal{A}=\frac{1}{4(8\pi G)^2T^2}+\frac{\sqrt[3]{\frac{128}{3}}(\frac{\alpha T}{2\pi G})^{\frac{2}{3}}}{\Big(9+\sqrt{3(27-256(8\pi G)^2(4\alpha)T^4)}\Big)^{\frac{1}{3}}}+\frac{\Big(\frac{4\alpha}{(8\pi G)^4T^2}\Big)^{\frac{1}{3}}\Big(9+\sqrt{3(27-256(8\pi G)^2(4\alpha)T^4)}\Big)^{\frac{1}{3}}}{\sqrt[3]{18}}.
\label{d8}    
\end{equation}
Notice that choosing the positive or negative sign affects the necessary conditions to have a positive mass. Moreover, the number of DOF is given by
\begin{equation}
N=4\Big[4\pi GM^2-\frac{\alpha\Big(1-(4\pi GM^2)^{1-q}\Big)}{1-q}+\beta\Big].
\label{d9}
\end{equation}
By inserting Eq.(\ref{d5}) into the above relation, we find the modified equipartition law by solving the main equation
\begin{equation}
\frac{M}{2T}+\frac{\alpha q}{1-q}(4\pi G)^{1-q}M^{2(1-q)}+\beta-\frac{N}{4}-\frac{\alpha}{1-q}=0,
\label{d10}    
\end{equation}
%or equivalently
%\begin{equation}
%\frac{M}{2T}+\alpha q\Big(\ln_q(4\pi G)+\frac{1}{1-q}\Big)M^{2(1-q)}+\beta-\frac{N}{4}-\frac{\alpha}{1-q}=0,
%\label{}    
%\end{equation}
that can be specified for different values of the non-extensivity parameter $q\neq1$. It is clear that the standard equipartition law $M=\frac{1}{2}NT$ is recovered for $q\rightarrow1$ and $\alpha=\beta=0$. Now, we attempt to solve the above equation for some values of the parameter $q$. For $q=1$, we trivially expect to approach the modified equipartition law obtained in the context of   logarithmic-corrected BH entropy shown in Eq.(\ref{b7}) or equivalently Eq.(\ref{b11}). For $q=2$, the standard equipartition law modifies as
\begin{equation}
M=\frac{2T}{3}\Big(\sqrt[3]{2}\frac{(\alpha+\beta-\frac{N}{4})^2}{\mathcal{B}}+\frac{\mathcal{B}}{\sqrt[3]{2}}-\alpha-\beta+\frac{N}{4}\Big),
\label{d11}    
\end{equation}
where
\begin{equation}
\mathcal{B}=\Bigg(-2(\alpha+\beta-\frac{N}{4})^3+\frac{27\alpha}{8\pi G T^2}+\frac{3}{2T}\sqrt{3\Big[-\frac{2\alpha}{\pi G}(\alpha+\beta-\frac{N}{4})^3+\frac{27\alpha^2}{16\pi^2G^2T^2}\Big]}\Bigg)^{1/3}.
\label{d12}    
\end{equation}
It is easy to see that the contribution of  black hole's DOF depends on the black hole temperature $T$ (via $\mathcal{B}$) in addition to the logarithmic-corrected BH entropy indices $\alpha$ and $\beta$ and the number of DOF $N$. Hence, adding non-extensivity effects from Tsallis statistics into the geometry of a logarithmic-corrected BH black hole destroys the validity of the equipartition theorem.
For the heat capacity of  black hole, we have 
\begin{equation}
C_V=-\frac{2(4\pi GM^2)^{1-q}\Big((4\pi GM^2)^{q}+\alpha\Big)^2}{(4\pi GM^2)^{q}+\alpha(1-2q)},
\label{d13}
\end{equation}
which reduces to the heat capacity of an unstable BH black hole in the case $q\rightarrow1$ and $\alpha=0$. The stability condition $C_V>0$ navigates us to consider the following constraint  
\begin{equation}
\frac{(4\pi GM^2)^{1-q}}{(4\pi GM^2)^{q}+\alpha(1-2q)}<0\hspace{0.5cm}\longrightarrow\hspace{0.5cm}
%\begin{cases}
    (4\pi GM^2)^{1-q}>0  \hspace{0.2cm} \text{and} \hspace{0.2cm} (4\pi GM^2)^{q}+\alpha(1-2q)<0,
    %\\(4\pi GM^2)^{1-q}<0  \hspace{0.2cm} \text{and}\hspace{0.2cm}  (4\pi GM^2)^{q}+\alpha(1-2q)>0.\end{cases}
  \label{d14}
\end{equation}
Although we can not present a clear judgment about the validity of the above condition, due to the lack of information about the relation between the model parameters, we find that the necessary condition (\ref{d14}) trivially reduces to the case of the unstable logarithmic-corrected BH black hole (\ref{b12}) when the non-extensivity parameter $q$ goes to the unit. For $q=2$, one can see that the condition (\ref{d14}) leads to the inequality $3\alpha>(4\pi GM^2)^2$, referring to a \textit{slightly} deviated version of the  logarithmic-corrected BH black hole ($q=1$), which is unstable even in  presence of the non-extensivity factor $q=2$. Now, we attempt to understand the effects of the non-extensivity index $q=2$ on the lifetime of the  logarithmic-corrected BH black hole by introducing the modified Boltzmann radiation law
\begin{equation}
J_\text{TLC($q=2$)}=-\frac{\frac{dE}{dt}}{\mathcal{A}}=\sigma T^4,
\label{d15}    
\end{equation}
where $\mathcal{A}=A_g+\alpha\ln_q(A_g)+\beta$ and $\sigma$ is the Stefan-Boltzmann constant. By inserting the black hole area and the temperature into the above relation, one  obtains 
\begin{equation}
\frac{dM}{dt}=-\frac{\sigma\Big(16\pi G^2M^2-\frac{\alpha}{16\pi G^2M^2}+\alpha+\beta\Big)}{\big(8\pi GM+\frac{\alpha}{2\pi GM^3}\big)^4}.
\label{d16}    
\end{equation}
By considering the instability constraint of the black hole and then neglecting the role of $\alpha$ with respect to the mass term, we find that the mass and lifetime of the  logarithmic-corrected BH black hole with a non-extensivity parameter $q=2$ vary almost similarly to their standard counterparts ($q=1$) shown in (\ref{b15}) and (\ref{b16}), respectively. Hence, studying higher values of $q$ provides a convenient framework to reveal the effects of the Tsallis statistics since the non-extensivity index $q=2$ \textit{minimally} contaminates the  logarithmic-corrected BH black surface. In addition to the above-discussed thermal properties, the Helmholtz free energy of the  logarithmic-corrected BH black hole, in the presence of the non-extensivity parameter $q$, can be studied using the formula $F=U-TS$. For $q=1$, we expect to reproduce the Helmholtz free energy of the  logarithmic-corrected BH black hole shown in (\ref{b17}), while, for $q=2$, we  find the corresponding Helmholtz free energy using the suitable expression of the black hole mass in terms of the temperature shown in Eqs. (\ref{d6}) and (\ref{d7}). 
%\begin{eqnarray}
%\Bigg\{-\frac{1024}{3}\pi^3G^3M^3-\frac{\alpha^3}{36\pi^3G^3M^9}+ 256\pi^2G^2M(\alpha+\beta)-\frac{\alpha^2(\alpha+\beta)}{7\pi^2G^2M^7}-\hspace{5cm}\nonumber\\
%-\frac{16\Big(\alpha^3+\beta^3+3\alpha^2 (2+\beta)+3\alpha\beta(2+\beta)\Big)}{3 M^3}-\frac{4\alpha(\alpha^2+\beta^2+\alpha (5+2\beta))}{5\pi GM^2}-\frac{64\pi G}{\alpha M}\times\hspace{2cm}\nonumber\\
%\times \Big(\alpha^4+\beta^4+\alpha^3(7+4\beta)+\alpha\beta^2(7+4\beta)+\alpha^2 (11+14\beta+6\beta^2)\Big)
%-\frac{64\sqrt{2}(\pi G)^{3/2}(\alpha^2+\beta^2+2\alpha(2+\beta))^{3/2}}{\alpha}\times\nonumber\\
%\Bigg(\frac{\alpha+\beta+\sqrt{\alpha^2+\beta^2+2\alpha(2+\beta)}}{\sqrt{\alpha+\beta-\sqrt{\alpha^2+\beta^2+2\alpha(2+\beta)}}}\arctan\Big[{\frac{2M\sqrt{2\pi G}}{\sqrt{\alpha+\beta-\sqrt{\alpha^2+\beta^2+2\alpha(2+\beta)}}}}\Big]+\hspace{2cm}\nonumber\\
%+\frac{\alpha+\beta-\sqrt{\alpha^2+\beta^2+2\alpha(2+\beta)}}{\alpha\sqrt{\alpha+\beta+\sqrt{\alpha^2+\beta^2+2\alpha(2+\beta)}}}\arctan{\frac{2M\sqrt{2\pi G}}{\sqrt{\alpha+\beta+\sqrt{\alpha^2+\beta^2+2\alpha(2+\beta)}}}}\Bigg)\Bigg\}\Bigg|^{M_{0}}_{M}=\sigma t.
%\label{}    
%\end{eqnarray}
\subsection{ The logarithmic-corrected Barrow black hole}
To study the effects of non-extensivity on the thermodynamic observables of a  logarithmic-corrected Barrow black hole, we start with the modified entropy
\begin{equation}
S_\text{TBLC}=\bigg(\frac{A_g}{A_{pl}}\bigg)^{1+\frac{\Delta}{2}}+\alpha(1+\frac{\Delta}{2})\ln_q\bigg(\frac{A_g}{A_{pl}}\bigg)+\beta,
\label{e1}    
\end{equation}
where the subscript $\text{TBLC}$ indicates the  logarithmic-corrected Barrow entropy with non-extensivity effects from the Tsallis statistics. Then, using the surface area of the black hole and Tsallis entropy (\ref{d2}), we can rewrite the entropy (\ref{e1}) as
\begin{equation}
S_{\text{TBLC}}=\big(4\pi GM^2\big)^{1+\frac{\Delta}{2}}-\alpha(1+\frac{\Delta}{2})\frac{\Big(1-(4\pi GM^2)^{1-q}\Big)}{1-q}+\beta,
\label{e2}
\end{equation}
which is related to the temperature 
\begin{equation}
T=\frac{M}{(2+\Delta)\Big((4\pi GM^2)^{1+\frac{\Delta}{2}}+\alpha(4\pi GM^2)^{1-q}\Big)}.
\label{e3}    
\end{equation}
Analogous to the previous section, we can find some analytical solutions for the mass of the black hole in terms of the temperature by solving the above equation in the case of different configurations of the parameters $q$ and $\Delta$. For the smooth factor $\Delta=0$, the standard case $q=1$ shows the result related to the  logarithmic-corrected BH model presented in (\ref{b4}), while the non-extensive case $q=2$ recovers the result of the previous section presented in Eqs.(\ref{d6}) and (\ref{d7}). For the maximal deformation $\Delta=1$ and by assuming $q=1$, we recover the results of the  logarithmic-corrected Barrow black hole shown in (\ref{cn1}), while the non-extensive case $q=2$ gives us the polynomial equation
\begin{equation}
(4\pi G)^{\frac{5}{2}}M^5-\frac{4\pi G}{3T}M^3+\alpha=0.
\label{e4}    
\end{equation}
It is clear that we are not able to present an analytical solution for the above equation since the Abel-Ruffini theorem says there is no general algebraic solution in radicals for polynomial equations of degree 5 or higher with arbitrary coefficients. Here, \textit{general} means that the coefficients of the equation are viewed and manipulated as indeterminates. Moreover, we find the following expression for the DOF 
\begin{equation}
N=4\Big[\big(4\pi GM^2\big)^{1+\frac{\Delta}{2}}-\alpha(1+\frac{\Delta}{2})\frac{\Big(1-(4\pi GM^2)^{1-q}\Big)}{1-q}+\beta\Big].
\label{e5}    
\end{equation}
Combining the two Eqs.(\ref{e3}) and (\ref{e5}), one can obtain 
\begin{equation}
\frac{M}{T(2+\Delta)}+\alpha\Big(-1+\frac{1+\frac{\Delta}{2}}{1-q}\Big)(4\pi G)^{1-q}M^{2(1-q)}+\beta-\frac{N}{4}-\frac{\alpha(1+\frac{\Delta}{2})}{1-q}=0,
\label{e6}    
\end{equation}
%or equivalently
%\begin{equation}
%\frac{M}{T(2+\Delta)}+\alpha (q+\frac{\Delta}{2})\Big(\ln_q(4\pi G)+\frac{1}{1-q}\Big)M^{2(1-q)}+\beta-\frac{N}{4}-\frac{\alpha(1+\frac{\Delta}{2})}{1-q}=0,
%\label{}    
%\end{equation}
that gives us the modified version of equipartition law. Note that, in the limit $\Delta=\alpha=\beta=0$ and $q\rightarrow1$, the standard equipartition law is recovered. Now, let us find the modified equipartition law related to different sets of the model parameters. For $\Delta=0$ and $q=1$, we expect to reach the modified equipartition laws (\ref{b7}) and (\ref{b11}) related to the  logarithmic-corrected BH black hole. For $\Delta=0$ and the non-extensive case $q=2$, we reproduce the equipartition law given in the previous section (\ref{d11}). For the most fractal structures with $\Delta=1$ and by considering $q=1$, the equipartition law reduces to the case of the  logarithmic-corrected Barrow entropy shown in (\ref{c6}) and (\ref{c8}). Moreover, for the maximal deformation case $\Delta=1$ with the non-extensive case $q=2$, the equipartition law is given by
\begin{equation}
M=T\Big(\sqrt[3]{2}\frac{(\frac{3\alpha}{2}+\beta-\frac{N}{4})^2}{\mathcal{B}}+\frac{\mathcal{B}}{\sqrt[3]{2}}-\frac{3\alpha}{2}-\beta+\frac{N}{4}\Big),
\label{e7}   
\end{equation}
where
\begin{equation}
\mathcal{B}=\Bigg(-2\Big(\frac{3\alpha}{2}+\beta-\frac{N}{4}\Big)^3+\frac{15\alpha}{8\pi G T^2}+\frac{1}{T}\sqrt{3\Big[-\frac{5\alpha}{2\pi G}\Big(\frac{3\alpha}{2}+\beta-\frac{N}{4}\Big)^3+\frac{75\alpha^2}{16\pi^2G^2T^2}\Big]}\Bigg)^{1/3}.
\label{e8}    
\end{equation}
Analogous to the previous case, inserting non-extensivity effects into the geometry of a logarithmic-corrected Barrow black hole, in the case of maximal deformation $\Delta=1$, perturbs the validity of the equipartition theorem due to the dependence of  black hole's DOF on the black hole temperature.
The heat capacity of the  logarithmic-corrected Barrow black hole, in the presence of the non-extensivity factor, is obtained as
\begin{equation}
C_V=-\frac{2(4\pi GM^2)^{1-q}\Big((4\pi GM^2)^{q}+\alpha\Big)\Big((4\pi GM^2)^{q+\frac{\Delta}{2}}+\alpha\Big)}{(1+\Delta)(4\pi GM^2)^{q+\frac{\Delta}{2}}+\alpha(1-2q)},
\label{e9}
\end{equation}
which reduces to the heat capacity of the BH black hole when $\alpha=\Delta=0$ and $q\rightarrow1$. It is easy to see that the stability condition $C_V>0$ leads to the constraint
\begin{equation}
\frac{\Big((4\pi GM^2)^{q}+\alpha\Big)\Big((4\pi GM^2)^{q+\frac{\Delta}{2}}+\alpha\Big)}{(1+\Delta)(4\pi GM^2)^{q+\frac{\Delta}{2}}+\alpha(1-2q)}<0.
  \label{e10}
\end{equation}
Analogous to the previous section, we  study the above condition for possible configurations of the parameters. For a black hole with a smooth surface $\Delta=0$, without any non-extensivity effect $q=1$, the condition (\ref{e10}) refers to the unstable  logarithmic-corrected BH black hole (\ref{b12}), while it reduces to Eq.(\ref{d14}) when the non-extensivity index $q=2$ is added to the black hole geometry. By considering the most fractal deformation $\Delta=1$ and in the absence of the Tsallis index $q=1$, the constraint (\ref{e10}) implies the unstable  logarithmic-corrected Barrow black hole (\ref{c10}). By taking into account the non-extensivity effect $q=2$ on the the black hole surface with maximal deformation $\Delta=1$, the stability constraint (\ref{e10}) reduces to $3\alpha>2(4\pi GM^2)^{\frac{5}{2}}$ when $\alpha>0$. By comparing this result with the stability condition of the  logarithmic-corrected Barrow black hole presented in (\ref{c10}), we find that the  logarithmic-corrected Barrow black hole with the non-extensivity index $q=2$ can be assumed a \textit{softly} deviated version of the unstable  logarithmic-corrected Barrow black hole $q=1$. Therefore, the  logarithmic-corrected Barrow black hole, in the context of the Tsallis statistics, is known as an unstable thermal system when its surface area, with the most fractal structure $\Delta=1$, is affected by the non-extensivity factor $q=2$.     

%\textcolor{cyan}{\section{Barrow entropy with relativistic effects}
%\section{A qualitative comparison between Barrow-based entropies}\label{s5}
\section{Discussion and Conclusions}\label{s5}
Black holes as gravitational systems can be understood in the realm of thermodynamics due to the description of their thermal properties in the context of the laws of thermodynamics. Also, the relation between gravity and thermodynamics is confirmed through the "gravity-thermodynamics" conjecture in which gravity is extractable from the laws of thermodynamics. The mentioned connection was proposed by Bekenstein and Hawking by introducing the well-known formula $S=A/4$ in which entropy, as a thermodynamic quantity, directly connects to the surface area of the black hole as a gravitational system. Also, Hawking discovered that, from the quantum mechanics viewpoint,  black holes emit thermal radiation through the Hawking radiation process. Hence, black holes interestingly provide a framework to make a connection between gravity, thermodynamics, and quantum physics. Despite the successes of the BH entropy, non-Gaussian statistics have been proposed to describe gravitational and cosmological situations in the presence of the non-extensive generalization of the horizon DOF or quantum gravitational corrections of the black hole surface. For instance, Tsallis statistics is a non-extensive generalization of the BG statistics characterized by the index $0\leq q<1$ and $q>1$ related to the rare and frequent events, respectively. Also, Kaniadakis proposed a relativistic generalization of the BG statistics, dealing with a non-exponential distribution function parameterized by the index $-1<\kappa<1$. In addition to the above statistics, considering quantum gravity effects on the black hole surface leads to defining some interesting generalizations of the BH statistics. A logarithmic-corrected version of the BH entropy occurs by taking into account self-gravity and backreaction effects in the tunneling formalism of the Hawking radiation. In the context of Barrow entropy, Barrow claimed that quantum gravitational
effects can produce some fractal structures on the Schwarzschild black hole surface tending to infinity while the volume is finite. These quantum effects, in the framework of Barrow entropy, are parameterized by the index $0\leq\Delta\leq1$ in which $\Delta=0$ and $\Delta=1$ depict the smooth and the most fractal structures, respectively. 

In this work, we  explored the thermal properties of a Schwarzschild black hole represented by different types of Barrow entropy. We first analyzed the issue from the viewpoint of the standard Barrow entropy. Then, we studied the thermodynamic quantities of the black hole in  presence of a logarithmic term, derived from LQG effects, added to the Barrow entropy. Moreover, we investigated the thermodynamics of a Barrow black hole when some non-extensivity effects, from the Tsallis statistics, are considered in the black hole geometry. Let us summarize the obtained results as follows:
\begin{itemize}
    \item \textbf{Barrow entropy:}\\
    From the viewpoint of the Barrow statistics, a Schwarzschild black hole is an unstable thermal system since the stability condition $-2<\Delta<-1$ is out of the theoretically allowed range of the Barrow index $0\leq\Delta\leq1$. Notwithstanding the instability of the black hole, the Barrow entropy extends the lifetime of the black hole by increasing the value of the Barrow index. Thus, in the presence of a non-zero Barrow index, the evaporation process of the Barrow black hole takes a longer time than the BH black hole ($\Delta=0$). Moreover, the entropy of the  Barrow black hole approaches zero slower than its counterpart in the BH black hole. In addition to the above results, we confirmed the instability of the Barrow black hole by calculating the Helmholtz free energy and representing its plot versus black hole temperature. Also, from the partition function of the Barrow black hole, we realized that the general microscopic features of the Schwarzschild black hole are preserved in the context of the Barrow statistics.
    \item \textbf{Logarithmic-corrected Barrow entropy:}\\
    Due to the inconsistency of the stability condition of a logarithmic-corrected Barrow black hole with the obtained constraint on the model parameters, the Barrow black hole is unstable even in  presence of self-gravity and backreaction effects on its geometry. In the  logarithmic-corrected Barrow statistics, the black hole evaporation process takes a longer time than its counterpart in the  logarithmic-corrected BH statistics. Hence, the black hole lifetime is extended due to the increasing  values of the Barrow index. Also, the entropy of the  logarithmic-corrected Barrow black hole declines slower than the  logarithmic-corrected BH black hole. Also, the Helmholtz free energy of both logarithmic-corrected BH and logarithmic-corrected Barrow black holes endorse the obtained results from their heat capacity, showing the instability of the Schwarzschild black hole. Analogous to the Barrow black hole, the partition function of both logarithmic-corrected BH and logarithmic-corrected Barrow black holes do not show any significant change in the microscopic behavior of a Schwarzschild black hole.         
    \item \textbf{Barrow entropy with non-extensivity effects:}\\
    The  logarithmic-corrected Barrow black hole with the most fractal structure $\Delta=1$,  affected by the non-extensivity factor $q=2$, can be considered a \textit{softly} deviated version of the unstable  logarithmic-corrected Barrow black hole with $q=1$. Also, the  logarithmic-corrected BH black hole ($\Delta=0$) in the context of the Tsallis statistics with non-extensivity factor $q=2$ behaves as a \textit{slightly} deviated version of the unstable  logarithmic-corrected BH black hole with $q=1$. Thus, the  logarithmic-corrected Barrow black hole, with the most fractal structure $\Delta=1$, and also the  logarithmic-corrected BH black hole ($\Delta=0$), in the context of the Tsallis statistics, are unstable when their surface areas are perturbed by the non-extensivity factor $q=2$.
\end{itemize}

The above-summarized results point out that considering power-law or logarithmic corrections, coming from LQG effects, to the BH entropy plays a crucial role in delaying the evaporation process of the Schwarzschild black hole due to the presence of non-zero indices related to the Barrow or BH logarithmic-corrected statistics. From the viewpoint of the  logarithmic-corrected Barrow statistics, where both power-law and logarithmic corrections are imposed on the BH entropy, the Schwarzschild black hole experiences even a longer evaporation process through non-zero statistics indices. By adding the non-extensivity effects, coming from the Tsallis statistics, to the geometry of the  logarithmic-corrected Barrow black hole, the results tell us that the black hole with the non-extensivity index $q=2$ behaves as a \textit{slightly} deviated version of the unstable  logarithmic-corrected Barrow black hole with $q=1$. Hence, by working with higher values of $q$, we can see how the thermodynamic quantities of the black hole are modified in the context of the Tsallis entropy.  
  
\section*{Acknowledgments}
 SC acknowledges the support of Istituto Nazionale di Fisica Nucleare, Sezione di Napoli,  Iniziative Specifiche QGSKY and MoonLight-2. This article is based upon work from COST Action CA21136 Addressing observational tensions in cosmology with systematic and fundamental physics (CosmoVerse) supported by COST (European Cooperation in Science and Technology).
 
\bibliographystyle{ieeetr}
\bibliography{biblo}
\end{document}